\documentstyle[twocolumn,psfig,fleqn,amstex,rotating]{mn}
\newcommand{\hi}{\mbox{H\ {\footnotesize I}}}

\def\etal{{et al.\ }}

\def\lsim{~\rlap{\raise 0.4ex\hbox{$<$}}{\lower 0.7ex\hbox{$\sim$}}~}
\def\gsim{~\rlap{\raise 0.4ex\hbox{$>$}}{\lower 0.7ex\hbox{$\sim$}}~}
\def\dd{{\rm d}}

\def\ppsi{{{\psi}}}

\def\tcbr{T_{_{\rm CBR}}}
\def\ts{T_{\rm s}}

\def\tb{T_{\rm b}}
\def\xhi{x_{_{\rm HI}}}
\def\xis{\xi^{\rm s}}
\def\xir{\xi^{\mathrm r}}

\def\txis{{\tilde \xi}^{\mathrm s}}
\def\tmu{ {\tilde \mu}}
\def\tr{ {\tilde r}}
\def\vr{{\bf r}}
\def\vs{{\bf s}}
\def\vv{{\bf v}}
\def\rpe{{\bf r}_{\perp}}
\def\rpa{{r}_{\parallel}}
\def\spe{{\bf s}_{\perp}}
\def\spa{{s}_{\parallel}}
\def\trpe{ {\tilde {\bf r}}_{\perp}}
\def\trpa{{\tilde r}_{\parallel}}
\def\vs{{\bf s}}

\def\l0{L_\ast(0)}

\def\s0{S_\ast(0)}

\def\omg0{\Omega_0}
\def\dd{{\mathrm d}}

\def\a2{\alpha^{(2)}}

\def\mpc3{\ {\rm Mpc^{-3}}}
\def\gpc3{\ {\rm Gpc^{-3}}}

\def\hmpc{\ {\rm h^{-1}Mpc}}

\def\kms{\ {\rm km\ s^{-1}}}

\begin{document}

\title[]
{The Alcock-Paczy\'nski test in redshifted twenty one centimeter maps}
\author[Nusser]{Adi Nusser\\\\
Physics Department and the Asher Space Researrch Institute, Technion, Haifa 32000, Israel\\}
\maketitle

\begin{abstract}
We examine the possibility of constraining the cosmological
 mean mass and dark energy densities 
by an application of the Alcock-Paczy\'nski test on redshifted 21-cm maps of the epoch of
reionization.
The 21-cm data will be provided
as a function of frequency and angular positions on the sky. 
The ratio of the frequency to angular distance scales
can be determined by  inspecting the anisotropy pattern of the  
correlation  function of brightness temperature.
We assess the sensitivity of the distance ratio to the cosmological
parameters and
 present a technique for disentangling geometric distortions 
from redshift space distortions caused by peculiar motions.
\end{abstract}



\section {Introduction}
\label{sec:introduction}

Observations of the cosmic microwave background (CMB)  and 
Type 1a Supernova place significant constraints on the
physical parameters  governing the evolution 
 of the cosmological background (Spergel \etal 2003; Knop \etal 2003;
 Riess \etal 2004).
The main parameters constrained by these observational data  are the mean matter (mass) density,
$\Omega_{\rm m}$, 
 and 
the density of negative pressure  energy (dark energy), $\Omega_{\rm v}$.
Perhaps the most intriguing  implication of the observations
is the need for a dark energy component which dominates the energy budget 
in the Universe today (Ratra \& Peebles 1988;  Wetterich 1995;  Coble, Dodelson \& Frieman 1997; Turner \& White 1997; Wang \etal 2000; Caldwell \& Doran 2004; Kunz \etal 2004; Caresia, Matarrese \&  Moscardini 2004).
By and large the constraints derived from these observations are
sustained 
by other estimates based on analyses 
of cosmic flows (Strauss \& Willick 1995; Nusser, Willick \& Davis 1997; Zaroubi \etal 2001),
abundance of rich galaxy clusters and 
its evolution (Bahcall \etal 2003; Ikebe \etal 2002; but see Vauclair \etal 2003) the mass-density power spectrum derived from 
galaxy redshift surveys (Percival \etal 2002; Zehavi \etal 2002; Tegmark \etal 2004), and the Ly-$\alpha$ forest (Croft \etal 2002, McDonald \etal 2000, Nusser \& Haehnelt 2000, McDonald \etal 2004,
Viel, Weller \& Haehnelt 2004).
Nevertheless all these  constraints are derived
from measurements at either very high  (the CMB) or low and intermediate redshifts (SN and other data), excluding 
the wide redshift range  the last scattering surface of CMB photons until $z\sim 4$.
Here we present in detail a technique for constraining the cosmological parameters
by an application of the Alcock-Paczy\'nski (AP) test  (Alcock \& Paczy\'nsky 1979;
see also Hui, Stebbins \& Burles 1999; Ballinger, Peacock \& Heavens 96, da \^Angela, Outram \& Shanks 2005) on future data at $z\sim 6-30$. These data are 
maps of the redshifted 21-cm emission/absorption lines 
produced by  \hi\ in the high redshift universe (Field 1959; Sunyaev \& Zel'dovich 1975; Hogan \& Rees 1979; Subramanian \&  Padmanabhan 1993; Madau, Meiksin \& Rees 1997). 
The 21-cm line is produced in the transition between the triplet and
singlet sublevels of the  hyperfinestructure of the ground level of neutral hydrogen atoms. 
A patch of \hi\  would be  visible  against CMB
when its spin temperature $\ts$ differs the CMB temperature, $\tcbr$.
Various mechanisms exist for raising $\ts$ significantly above 
$\tcbr$ during the era of reionization.
As a result a significant cosmological signal from the era of reionization  should be expected (Field 1958 \& 1959;  Scott \& Rees 1990; Madau, Meiksin \& Rees 1997;  
 Baltz, Gnedin \& Silk 1998; Tozzi \etal 2000; Ciardi \& Madau 2003; Chen \& Miralda-Escude 2004; Gnedin \& Shaver 2004; Nusser 2005; Ricotti, Ostriker \& Gnedin 2004).

We advocate the 
the application of the AP test  on the correlations
measured in three dimensional maps of 21-cm emission. 
These correlations are readily expressed in terms of angular separations and 
frequency intervals. In order to derive the correlations in terms of real space separations (measured for example in $\hmpc$) one needs to know 
the mean density parameter, $\Omega_{\rm m}$, and the dark energy 
density $\Omega_{\rm v}$ and its temporal evolution (or equivalently 
its equation of state). Constraints on these  parameters can then be  
obtained by demanding isotropy of correlations in real space.
There are two main obstacles in implementing this program on future 
data. The first is foreground contamination 
(e.g. Oh \& Mack 
2003; Di Matteo, Ciardi \& Miniati 2004) which we will not discuss here.
The second is redshift distortions.
The redshift coordinate of a patch of matter is given 
by the sum of its true  real space coordinate and the 
its line of sight peculiar velocity (deviation from Hubble flow).
The distribution of matter in redshift space should therefore
be different from the real space distribution.
According to the standard cosmological paradigm, the observed
large scale structure have formed by gravitational amplification 
of tiny initial fluctuations (Peebes 1980, Peacock 1999). 
In late time  linear theory the  
gravitational growth of mass density fluctuations
is inevitable associated with 
coherent peculiar motions. Therefore, over large scales where 
linear theory is generally valid the density in redshift space
appears more clustered than in real space. Further, because
because redshift and real space coordinates differ only in the line
of sight direction, the structure in redshift space would appear anisotropic.
On small scales random motions  are important, causing a suppression of clustering
in redshift space. Here we  restrict the
analysis to linear theory which is valid at the high redshifts ($z\sim 10-30$) and 
the physical scales ($\gsim 1\hmpc$) we  consider. 
Redshift distortions, in the linear as well the non-linear regime, have been investigated in great detail in relation to the
galaxy redshift surves (Peebles 1980; Davis \& Peebles 1983; Kaiser 1987; Hamilton 1992; Cole, Fisher \& Weinberg 1994; Nusser \& Davis 1994, Fisher \& Nusser 1996; Taylor \& Hamilton 1996; Zaroubi \& Hoffman 1996; Hatton \& Cole 1998; Desjacques \etal 2004) and to a lesser extent 
in relation  to 21-cm maps (Illiev \etal 2002; Furlanetto, Sokasian \& Hernquist
2004; Barkana \& Loeb 2004, Bharadwaj \& Saiyed 2004)). 

 In  this paper we show that the pattern of redshift space anisotropies
is quite distinct  from the geometric distortions introduced  by an erroneous
choice of the cosmological parameters. Further, we show that an application 
of the AP  test  will provide useful
constraints on the cosmological parameters if the correlation function 
of the cosmological 21-cm brightness temperature could be estimated 
to within a $20\%$ accuracy.

The outline of the paper is as follows. In \S\ref{s:bt}
 the relation between   21-cm emission/absorption and the
gas density is briefly summarized.  
\S\ref{s:zdist} describes in some detail redshift distortions in the 
linear regime and the expected anisotropy the density correlation 
function in redshift space.  
In \S\ref{s:gdist} the dependence of geometric distortions on 
the cosmological parameters is explored. 
\S\ref{s:tdist} shows how to distinguish between geometric and 
redshift space distortions. 
In \S\ref{s:alternative} we discuss additional 
application of the AP test with 21 cm maps.
We conclude with a short discussion in 
\S\ref{s:disc}.

\section{The brightness temperature}
\label{s:bt}
Intensities, $I(\nu)$,  at radio frequency are expressed in terms of brightness temperature
defined as $\tb=I(\nu)c^{2}/2k\nu^{2}$, where $c $ is the speed of light and $k$ is Boltzman's constant (Wild 1952).
The {\it differential brightness temperature} (hereafter, DBT) against the CMB
of a small patch of gas with spin temperature $\ts$ at redshift $z$ is (e.g. Ciardi \& Madau 2003),
\begin{eqnarray}
\nonumber \delta \tb&=&16{\rm \; mK} \; x_{_{\rm HI}}(1+\delta)\left(1-\frac{\tcbr}{\ts}\right)\\
&&\times 
\left(\frac{\Omega_{\rm b} h}{0.02}\right) 
\left[  \left( \frac{1+z}{10} \right) \left(\frac{0.3}{\Omega_{\rm m}}\right)
\right]^{1/2}\; . 
\label{tb}
\end{eqnarray}
where    $x_{_{\rm HI}}$ are 
is the fraction of \hi\ in the patch. The quantity $\delta=\rho/{\bar \rho}-1$ is the density contrast of the gas.
The fraction    $x_{_{\rm HI}}$ may depend on position, $\vr$, and  the density contrast, $\delta$ (e.g. Knox \etal 1998, Buscoli \etal 2000, Benson \etal 2001). 
This dependence is mainly dictated by the nature of the ionizing radiation
and the spatial distribution of its sources.
If the early staged of reionization are dominated by  X-ray photons
which have very large mean free path then $x_{_{\rm HI}}$ at any point in 
space is determined by the equilibrium between photoionization and recombinations
(e.g. Ricotti \& Ostriker 2004). In contrast,  UV photons from stellar sources
have a short mean free path. Therefore, in the presence of UV photons alone rionization proceeds in a patchy way with $x_{_{\rm HI}}$ 
very close to zero in ionized regions and unity otherwise.
We  model reionization in the presence of X-rays and UV photons  as follows.
We assume that the explicit  dependence of $x_{_{\rm HI}}$ on $\delta$ outside 
the regions ionized by UV is linear 
so that,
\begin{equation}
x_{_{\rm HI}}(\vr,\delta)\propto {\cal M}({\vr})\left[1+\frac{\partial \ln x_{_{\rm HI}}}{\partial \delta}\; \delta\right] \; ,
\label{eq:mask}
\end{equation} 
where $\cal M(\vr)$ is termed the mask and is either zero or unity. 
The expansion of $x_{_{\rm HI}}$ to first order in $\delta$ is justified here since we only 
consider   correlations on  scales where the density contrast is small.

In \S\ref{s:tdist} we will present a formalism for an application of the AP
test using the correlations of $\delta \tb$. In developing that formalism we
will assume that the mask structure (correlation) function is insensitive to 
peculiar velocities  and also  
uncorrelated with the 21 cm signal. We will see that these conditions
are valid in the limit of large filling factors ($F_{\rm f}\gsim 0.5$) of ionized regions 
At the early stages of re-ionization, when the filling factor
  is small enough,
 it will be possible to identify large unionized regions 
 in  future maps of 21 cm (see figure 3 of Ciardi \& Madau 2003). At these early stages, all complications related to the mask, $\cal M$, can 
 completely be avoided  by restricting the analysis to these regions. 
 In an x-ray {\it pre-reionization} scenario, X-rays 
 from accreting black holes contribute to partial ionization before 
 the appearance of significant UV sources (Ricotti \& Ostriker 2004). 
 During this phase, the mask is  unity except in the immediate  vicinities of the X-ray emitting sources (Zaroubi \& Silk 2005). The number density of X-ray sources
 is expected to be  low (Ricotti \& Ostriker 2004) so that role of the mask
 is negligible in this case.

\section{redshift distortions}
\label{s:zdist}
For simplicity of discussion we consider a 
region in the shape of a cubic box at redshift $z$.
Further, we assume the 
``distant observer limit'' (e.g. Kaiser 1987, Zaroubi \& Hoffman 1996) according 
to  which 
 the box is small compared to its distance from the observer.
We   work with a Cartesian coordinate system defined 
by the axes of the box  assuming that one of these axes
is in the direction of the line of sight to the box.
The box is assumed to be comoving with the Hubble flow 
and large enough so that the center of mass of matter inside it is
also comoving with the Hubble flow.

In the coordinate system attached to the box, let $\vv$ and $\vr$ be, repsectively, the physical peculiar velocity (deviations
from Hubble flow) and physical real space coordinate (i.e. Eulerian)
of a particle, both expressed in $\kms$. The redshift space coordinate 
is defined as 
$\vs=\vr + v_\parallel\hat {\bf l}$, where  
${\hat {\bf l}}$ is a unit vector in the line of sight direction and 
$v_\parallel=({\hat {\bf l}}\cdot \vv)$ is the line of sight peculiar velocity\footnote{The 
observed redshift coordinate  also contains  contributions from
the cosmological expansion and  from the preculiar motion of the observer. 
However, these contributions do not amount to any systematic effect.}. 
It convenient to denote  
the components of $\vr$ parallel and perpendicular to the line of sight 
by  $\rpa$ and $\rpe$, respectively. Using similar notation 
for $\vs$ we find $\spe=\rpe$ and $\spa=\rpa+v_{\parallel}$.

Only in the absence of peculiar motions (both thermal and cosmological)  
the density contrast,  $\delta$, appearing in (\ref{tb}) refers to the
actual real space density constrast, $\delta^{\rm r}$. 
In the presence of peculiar motions $\delta$ is  equal to the gas density in redshift space,  $\delta^{\rm s}$. This is given by
\begin{eqnarray}
\nonumber
&&1+\delta^{\rm s}(\spe,\spa)=\\
&&\int\dd \rpa \left[1+\delta^{\rm r}(\rpe,\rpa)\right]
\; {\rm G}\left[\rpa+U(\rpe,\rpa)-\spa\right] \; ,
\end{eqnarray}
where  we have  decomposed the line of sight peculiar velocity into a smooth component
$U(\vr)$ and a random thermal component, $v_{\rm th}$ having  a gaussian probability distribution 
function, ${\rm G}(v_{\rm th})$, 
of mean zero and rms $\sigma_{\rm th}$.
In the following we assume that $\sigma_{\rm th}$ is negligible so that  
$G$ reduces to a Dirac delta function. In this case we
obtain 
\begin{equation}
1+\delta^{\rm s}(\vs)=\left[1+\delta^{r}(\vr )\right] \left(1-\frac{\dd \rpa}{\dd \spa}\right)
\; ,
\label{eq:zden}
\end{equation}
where $\rpa$ is determined by $\rpa+U(\rpe,\rpa)=\spa$. 
Expanding (\ref{eq:zden}) to  first order in $\delta $ and $U$ yields
\begin{equation}
\delta^{\rm s}(\vr)=\delta^{\rm r}(\vr)-\frac{\dd U}{\dd \rpa}\; .
\label{eq:zdenlin}
\end{equation}
Note that to first order this last relation could equally be expressed
as a function of $\vs$ rather than $\vr$ (Nusser \& Davis 1994).
The term $\dd U /\dd \rpa$ introduces anisotropies in the 
density field in redshift space. 
Supplemented with a relation between the  velocity and density fields of the gas,
the relation (\ref{eq:zdenlin}) allows us to 
determine $\delta^{\rm s}$ from $\delta^{\rm r}$.
On large scale, a density-velocity relation 
between the real space mass density and velocity
fields can be obtained using linear perturbation theory  (e.g. Peebles 1980; 
Peacock 1999) .
This relation reads
\begin{equation}
\delta^{\rm r}=-\Omega_{\rm m}^{-0.6}{\bf \nabla} \cdot \vv \; .
\label{eq:vdlin}
\end{equation}
As an illustration of redshift distortions we show in figure 
(\ref{fig:map}) slices of the density fields in real and redshift spaces in the
bottom left and top left columns, respectively.
The real space density field is a gaussian random field generated with   
the  linear density power spectrum of the cold dark matter cosmology (CDM)
with  $\Omega_{\rm m}=0.3$. 
The linear relations (\ref{eq:zdenlin}) and (\ref{eq:vdlin}) have been used to derive 
$\delta^{\rm s}$ for a  line of sight is in the y-direction. 
The density fields are shown in a slice of $128\hmpc$ on the side.
To improve the visual presentations the fields have been 
smoothed with a   gaussian window of $3\hmpc$ width. 
The normalization is arbitrary and the contour spacing is 2.
 A comparison between the top left and bottom left panels 
 reveals significant differences between  $\delta^{\rm s}$ and $\delta^{\rm r }$.
The lack of isotropy in the contour maps of $\delta^{\rm s}$ is also clear 
(see Kaiser 1987 for more details).

The relation (\ref{eq:vdlin}) can easily be generalized to account for 
any {\it bias} between $\delta^{\rm r}$ and the actual mass density,
$\delta_{\rm m}^{\rm r}$. To first order in the density contrast such
a bias can be described  
by $\delta^{\rm r}=
b \delta^{\rm r}_{\rm m}$ where $b$ is a constant.
Then the relation (\ref{eq:vdlin}) is modified 
by multiplying the right hand side by $b$. 
Hamilton (1992) used the linear relations (\ref{eq:zdenlin}) and (\ref{eq:vdlin}) in order 
 to express the redshift space density correlation
function $\xis=<\delta^{\rm s}(\vr_{0})\delta^{\rm s}(\vr_{0}+\vr>_{\vr0}$ 
in terms of the real space correlation $\xir=<\delta^{\rm r}(\vr_{0})\delta^{\rm r}(\vr_{0}+\vr>_{\vr0}$
 as follows,
\begin{equation}
\xis (\rpa,\rpe)=
\xi_{0}(r)P_{0}(\mu) +\xi_{2}(r)P_{2}(\mu)+\xi_{4}(r)P_{4}(\mu) \; ,
\label{eq:corrzr}
\end{equation}
where $\rpa$ and $\trpe$ are the components of $\vr$ parallel and perpendicular to the line of sight, $\mu=\rpa/r$ is cosine the angle between $\vr$ and the line of sight, and
$P_{l}$ is the $l^{\rm th}$ order Legender polynomial: $P_{0}=1$, $P_{2}=(3\mu^{2}-1)/2$,
 $P_{4}=(35\mu^{4}-30\mu^{2}+3)/8$, and
 $P_{6}=(231 \mu^{6}-316 \mu^{4}+105 \mu^{2}-5)/16$ (we list $P_{6}$ for 
 subsequent use). Further, 
\begin{eqnarray}
\xi_{0}(r)=\left(1+\frac{2}{3}+\frac{1}{5} \beta^{2}\right) \xir(r)\\
\xi_{2}(r)=\left(\frac{4}{3}\beta+\frac{4}{7}\beta^{3}\right) \left[\xir(r)-
{\bar {\xi^{\rm r}}}\right]\\
\xi_{4}(r)=\frac{8}{35}\beta^{2} \left[\xir(r) +\frac{5}{2} {\bar {\xi^{\rm r}}}-
\frac{7}{2} {\bar {\bar {\xi^{\rm r}}}}\right]
\end{eqnarray}
where $\beta=\Omega_{\rm m }^{0.6}/b$, 
$ \bar {\xi^{\rm r}} =3r^{-3}\int_{0}^{r}\xir(s) s^{2}\dd s $,  
and $\bar {\bar {\xi^{\rm r}}} =5r^{-5}\int_{0}^{r}\xir(s) s^{4}\dd s $.

\section{geometric distortions}
\label{s:gdist}
\begin{figure}
\centering
\mbox{\psfig{figure=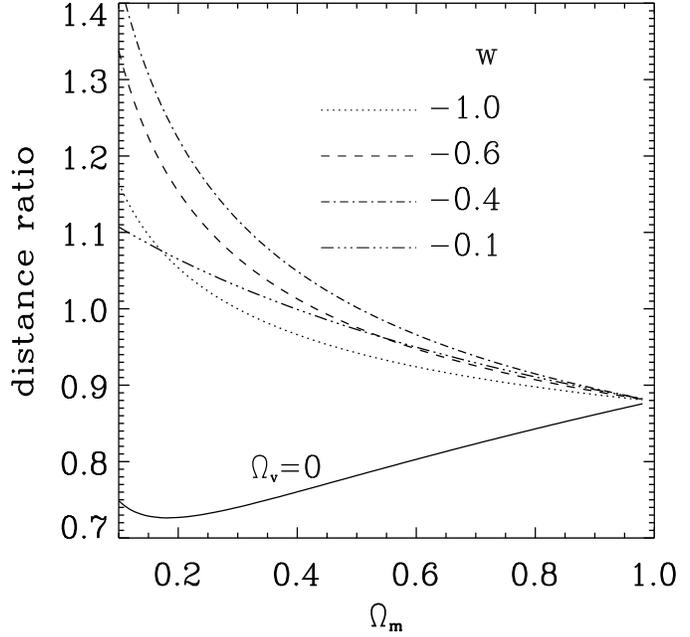,height=3.5in,width=3.5in}}
\caption{
The dependence  of the distance ratio $1/(H D_{\rm A})$ at $z=20$  
on the assumed cosmological parameters. The ratio is normalized to
its value at $\Omega_{\rm m}=0.3$, $\Omega_{\rm v}=0.7$,
 and $w=-1$. The solid curve is for an open universe  with matter 
content only. All other curves correspond to  universes with $\Omega_{\rm m}+
\Omega_{\rm v}=1$ for various values of $w$, as indicated in the figure.}
\label{fig:dom}
\end{figure}

\begin{figure}
\centering
\mbox{\psfig{figure=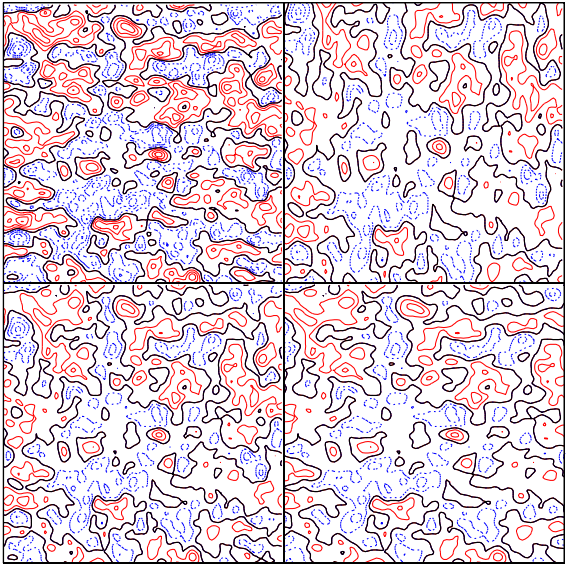,height=3.5in,width=3.5in}}
\caption{
An illustration of the redshift and geometric distortions.
The bottom left panel with the isotropic contours shows the density field in real space. The 
top left is the corresponding redshift space density field obtained
from the linear relations (\ref{eq:zdenlin}) and (\ref{eq:vdlin}) for 
a line
of sight  along the vertical axis. The top right and the bottom right panels
illustrate geometric distortions parallel and perpendicular the line
of sight, respectively. These geometric 
distortions are for a 25\%  difference between the assumed and true
distances 25\%.}
\label{fig:map}
\end{figure}

Twenty one centimeter  maps will be  readily provided 
in terms of the frequency along the line of sight direction and 
angular positions on the sky in the perpendicular plane. 
An observed frequency interval $\Delta \nu$ corresponds to 
a physical separation (in $\kms$), $\Delta \rpa$, of
\begin{equation}
\Delta \rpa=\frac{\Delta \nu}{ 1420{\rm Hz}} \frac{c}{1+z} \; .
\end{equation}
This relation is independent of the cosmological parameters.
The projected separation (also in $\kms$), $\Delta \rpe$, corresponding 
to an angular separation of $\Delta \theta$ is 
\begin{equation}
\Delta \rpe=\Delta \theta H D_{\rm A}(z) \; ,
\end{equation}
 where $D_{\rm A}=D_{\rm A}(\Omega_{m},\Omega_{\rm v},z)$ is the 
 angular diameter distance and $H(\Omega_{m}, \Omega_{\rm v},z)$  is the 
 Hubble function evaluated at redshift $z$.
 Therefore, for a given angular separation one needs
 to know the cosmological parameters in order to determine $\Delta \rpe$.
 The dependence $H D_{\rm A}$ on the cosmological parameters
is illustrated in  figure (\ref{fig:dom}) which shows the quantity $1/(H D_{\rm A})$
 for  cosmological models with and without 
 a dark energy component (Lima \& Alcaniz 2000). We assume  a dark energy  equation of state 
  of the form $P_{\rm v}=w \rho_{\rm v}$, where $P_{\rm v}$ and $\rho_{\rm v}$ are, respectively, the pressure and density of the 
 dark energy component, and $w$ is  constant.
This equation of state implies $\rho_{\rm v}\propto (1+z)^{3(1+w)}$ and
so  $w=-1$ describes a cosmological constant (constant $\rho_{\rm v}$).  
 We do not consider here $w<-1$.
The figure shows  $1/(H D_{\rm A})$ normalized to its value\footnote{In this paper the numerical 
values of $\Omega_{\rm m}$ and $\Omega_{\rm v}$ 
always correspond to $z=0$.} for $\Omega_{\rm m}=0.3$,
$\Omega_{\rm v}=0.7$ and $w=-1$.  The solid line in the figure is for an open universe universe  with 
  $\Omega_{\rm v}=0$. 
 All remaining curves  are for a flat universe ($\Omega_{\rm m}+\Omega_{\rm v}=1$)
 and they correspond to several values of $w$ as indicated in the figure.
The distance ratio is clearly sensitive to the assumed cosmological parameters.
For the models plotted in the figure the  maximal difference is  about 40\% at 
$\Omega_{\rm m}=0.2$.
The panels to the right of figure (\ref{fig:map})  illustrate the nature of the geometric distortions. 
The top (bottom) panel  
show   the density contours obtained by stretching the 
vertical (horizontal) axis of the slice in the left bottom panel $25\%$. 
The distinction between geometric and redshift space distortions is
readily seen by comparing these maps the density field
in redshift space shown in the left top panel. 
Redshift distortions amplify the  
density fluctuations, while geometric distortions merely stretch the contours.

In the absence of peculiar motions and foreground contamination 
the correlation function of 
 the brightness temperature is isotropic for the true  $\Omega_{\rm m}$,
 $\Omega_{\rm v}$, and $w$. 
  However, we expect redshift distortions  to be present 
 as density fluctuations are associated with peculiar motions under the
 action of gravity. 
 We show in the next section  how to disentangle the geometric from redshift distortions, allowing 
 an  estimation of the cosmological parameters
 from the isotropy pattern of the correlation function.

 In  the remainder of this section we derive  several relations for later use.  
The relation between redshift and real space correlations
is invariant under  an isotropic scaling transformation of the form $ \vr \rightarrow 
g \vr$, where $g$ is a constant. 
Therefore, without loss of generality the effect of an incorrect 
choice of the cosmological parameters can be described 
by a transformation of the type $\trpa=\rpa/(1+\alpha)$ and ${\tilde \vr}_{\perp}=\vr_{\perp}$, where $(1+\alpha)$ is the ratio of the true
value of $H D_{\rm A}$ to the assumed value. 
For clarity of notation, we place  a tilde  over  functions defined in the space of 
$\trpa$ and $\trpe$.
If $f=f(\rpa,\rpe)$  then 
\begin{equation}
{\tilde f}(\trpa,\trpe)\equiv f[(\rpa, \rpe)] \; ,
\label{eq:ftilde}
\end{equation}
We can easily verify that the correlation function of $d$ computed in 
the space of $(\trpa,\trpe) $ is 
\begin{equation}
\txis (\trpa, \trpe)=\xis[(1+\alpha)\trpa,
\trpe]\; .
\end{equation}
According to (\ref{eq:ftilde}), 
if $f=f(r)$ is independent of the direction of $\vr$, then 
\begin{equation}
{\tilde f}\left(\trpa,\trpe\right)=f(r)
\end{equation}
where $r=\sqrt{(1+\alpha)^{2}\trpa^{2}+\trpe^{2}}=
\tr \sqrt{(1+\alpha)^{2}\tmu^{2}+(1-\tmu^{2})}$.
To  first order in $\alpha$ we find 
\begin{equation}
{\tilde f}\left(\trpa,\trpe\right)=f(\tr)\left[
1+\alpha\tmu^{2} \frac{\dd \ln f(\tr)}{\dd \ln \tr}
\right] \; \label{eq:alphaf} .
\end{equation} 
We will also need  the first order (in $\alpha$) relation 
\begin{eqnarray}
\mu^{2}={\tilde \mu}^{2}+2\alpha {\tilde \mu}^{2} (1-{\tilde \mu}^{2})\; .
\label{eq:alphamu}
\end{eqnarray}
between $\tmu=\trpa/\tr$ and $\mu=\rpa/r$.

\section{distortions in twenty one cm  maps}
\label{s:tdist}
We assume that $\ts\gg \tcbr $ so that $\delta \tb\propto \xhi(1+\delta^{\rm s})$ up to 
a multiplicative constant. 
We further substitute  the expansion (\ref{eq:mask}) for $\xhi$ and write 
 $\xhi(1+\delta^{\rm s})\propto {\cal M}({\vr}) [1+(b-1)\delta^{\rm r}+\delta^{\rm s}] $ to first order in $\delta$, where $b=1+\partial \ln \xhi /\partial \delta$. Note that $\xhi$  is determined by 
 the real space density $\delta^{\rm r}$, hence the appearance of $\delta^{\rm r}$.
For simplicity we will assume that the data yields the quantity 
\begin{equation}
d(\vr)={\cal M}(\vr)(1+\Delta^{\rm s})\; ,
\label{eq:data}
\end{equation}
where $\Delta^{\rm s}=(b-1)\delta^{\rm r}+\delta^{\rm s}$.
Using the linear relation (\ref{eq:zdenlin}) we find that $\Delta^{\rm s}=(b-1)\delta^{\rm r}+\delta^{\rm s}=b\delta^{\rm r}-\dd U/\dd r$ which   can be identified  as the redshift space density  contrast corresponding to the 
real space density field  $\Delta^{\rm r}=b \delta^{\rm r}$. 

\begin{figure}
\centering
\mbox{\psfig{figure=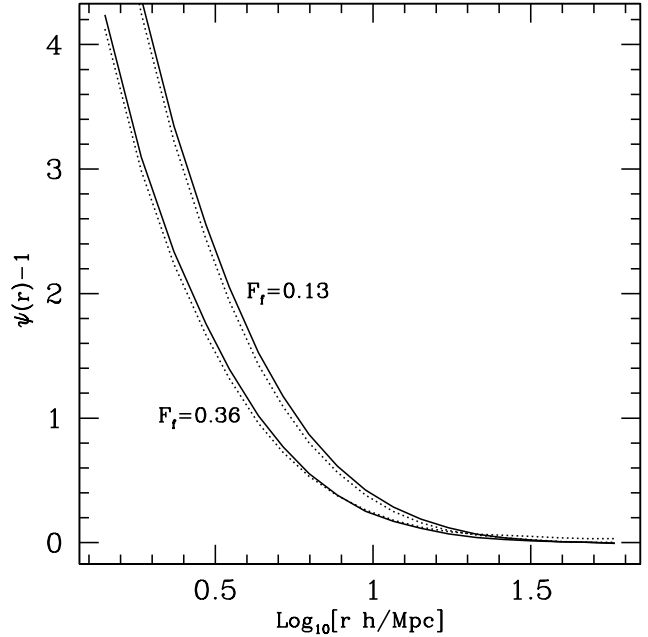,height=3.5in,width=3.5in}}
\caption{
The correlation  $\psi$ as a function of 
separation, for two values of the filling factor of ionized regions, as indicated in the figure.
The upper ($F_{\rm f}=0.13$) and lower sets ($F_{\rm f}=0.36$) of curves are for sources identified as 4$\sigma$ and
and $3.5\sigma$ peaks in the density field, respectively. 
The solid lines show the correlation  when the  mask of ionized regions is placed 
around the peaks, while  the dotted lines are for  masks  displaced
away from the peaks. }
\label{fig:psi}
\end{figure}

We write the data-data redshift space correlation function as
\begin{equation}
\psi^{\rm s}=<d d>={\cal C}_{_{\cal M}}(r)[1+\xis(\vr)] \; , 
\label{eq:corrmask}
\end{equation}
where ${\cal C}_{_{\cal M}}=<\cal M M>$ is the structure (correlation) function of the mask, and 
$\xis$ is now the correlation function of $\Delta^{\rm s}$. We have also assumed
that the mask $\xhi$ and $\delta$ are uncorrelated.
This assumption is justified once the filling factor of ionized regions
is larger than $\bar n R_{\rm c}^{3}$ where $\bar n$ is the number density of 
sources and $R_{\rm c}\sim \hmpc$ (comoving) is the coherence length of the correlations between 
the sources and the gas. This condition is easily satisfied as a single source 
is capable of ionizing relatively  large regions around it.
To test this assumption we have resorted to our   random gaussian 
density field having a $\Lambda$CDM power spectrum (see figure \ref{fig:map}).
We identify the sources as peaks in the density smoothed with $1\hmpc$ 
and assume that each source ionizes a region of a given radius around it. 
These regions define our mask. 
A comparison between the correlation of the mask times the density field 
when the mask is placed around the sources and 
when it is placed  randomly in the box serves as an indication to the 
the degree of correlation between the
mask and the density field.
In figure (\ref{fig:psi})  the solid lines 
are the correlation function $\psi$ (in real space)
for a mask placed around the sources while the dotted are the correlations when 
the mask is placed randomly in the box.
The upper and lower sets of curves correspond to two values of the filling factor, 
$F_{\rm f}$, of ionized regions.
The values $F_{\rm f}=0.13$ and $0.36$, respectively,  correspond to sources
identified as $4\sigma$ and $3.5\sigma$ peaks.  
The differences between the dotted and solid curves for each value of the $F_{\rm f}$ are negligible.
 
We will write now the expressions describing the 
anisotropy of the data-data correlation function 
in the presence of geometric and redshift distortions.
We restrict the analysis to the linear regime $\delta\ll 1$ and 
to first order in $\alpha$.

We write $\tilde \xis(\trpa,\trpe)=\xis( (1+\alpha)\trpa,\trpe)$ 
and use (\ref{eq:corrzr}) to express  $\xis $ in terms of the moments   $\xi_{l}(r)$
 with ($r=\sqrt{(1+\alpha)^{2}\trpa^{2}+\trpe^{2}} $) and Legendre polynomials evaluated 
at $\mu=(1+\alpha)\trpa/\tr$. Then we 
use  (\ref{eq:alphaf}) and (\ref{eq:alphamu}) to expand 
 $\xi_{l}$ and $P_{l}$ as functions of $\tr$ and $\tmu$ to first
 order in $\alpha$.
 The final result of this straightforward algebra is 
\begin{eqnarray}
\nonumber
&&{\tilde \psi}^{\rm s}\left(\trpa,\trpe\right)=
\displaystyle\sum_{l=0,2,4} \ppsi_{l}({\tilde r})P_{l}(\tmu)\\
\nonumber&&+\alpha \displaystyle \sum_{l=0,2,4}
\ppsi_{l}({\tilde r})\left[\frac{\dd \ln \ppsi_{l}({\tilde r})}{\dd \ln \tilde {r}} \tmu^{2}P_{l}(\tmu)
+ \tmu(1-\tmu^{2})\frac{\dd P_{l}(\tmu)}{\dd \tmu}\right] \\
&&=\displaystyle\displaystyle\sum_{l=0,2,4} \left[\ppsi_{l}({\tilde r}) +\alpha 
a_{l}(\tr) \right]P_{l}(\tmu)+\alpha a_{6}(\tr)P_{6}(\tmu) \; \,
\label{eq:gzdist}
\end{eqnarray}
\def\dfzero{\frac{\dd \ln \psi_{_0}}{\dd \ln {\tilde r}}}
\def\dftwo{\frac{\dd \ln \psi_{_2}}{\dd \ln {\tilde r}}}
\def\dffour{\frac{\dd \ln \psi_{_4}}{\dd \ln {\tilde r}}}
where $ \psi_{_{0}}=    {\cal C}_{_{\cal M}} (1+ \xi_{_{0}})$ and $\psi_{_{l}}= {\cal C}_{_{\cal M}} \xi_{_{l}}$ for $l>0$, and
\begin{eqnarray}
\nonumber
\nonumber
a_{0}&=&\frac{1}{3}\psi_{_0}\dfzero+\psi_{_2}\left(\frac{2}{5}+\frac{2}{15}\dftwo\right) \\
\nonumber
a_{2}&=&\frac{2}{3}\psi_{_{0}}\dfzero+\psi_{_{2}}\left(\frac{2}{9}+\frac{11}{21}\dftwo\right)
\\
\nonumber
&+&\psi_{_{4}} \left(\frac{20}{21}+\frac{4}{21}\dffour \right)\\
\nonumber
a_{4}&=&
-\psi_{_{2}} \left(\frac{24}{35}-\frac{12}{35}\dftwo\right)+
\psi_{_{4}} \left(\frac{20}{77}+\frac{39}{77}
\dffour\right) \\
a_{6}&=&-\psi_{_{4}}\left(\frac{40}{33}-\frac{10}{33}\dffour \right) \; .
\end{eqnarray}

\begin{figure}
\centering
\mbox{\psfig{figure=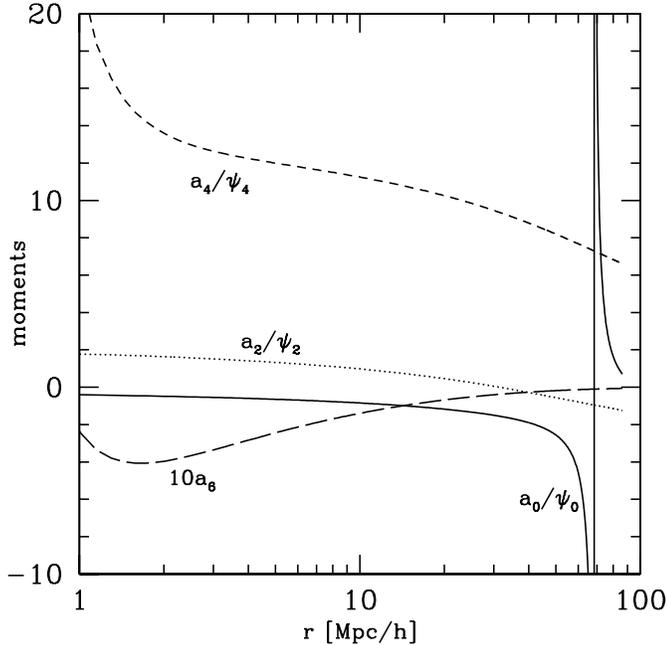,height=3.5in,width=3.5in}}
\caption{
Ratios of the moments, $a_{l}/\psi_{l}$ ($l=0,2,4$),  as a function of separation. 
Also shown is the moment $a_{6}$ multiplied by a factor of 10.
All curves correspond to the linear correlation function of $\Lambda$CDM with 
$\Omega_{\rm m}=0.3$, assuming that the  correlation of the mask is unity.}
\label{fig:moments}
\end{figure}

The visual impression between geometric and redshift distortions 
seen in figure (\ref{fig:map}) is quantified by the relation (\ref{eq:gzdist}). 
The moments $a_{l}$ and $\psi_{l}$, respectively,  characterize  the 
geometric and redshift distortions. A term containing $P_{6}$ is missing from 
the  redshift distortions of the correlations. One could  
determine the correct value of the distance ratio, $\alpha$, by
 minimizing the square of the moment corresponding to  $P_{6}$, independent of 
 the details of the mask structure function ${\cal C}_{_{\cal M}}$.
 As we shall see below other moments can also be used to infer $\alpha$. 
Let us inspect 
 the moments $a_{l}$ and $\psi_{l}$. 
The moments depend on the linear matter correlation function which we obtain 
 from the power spectrum of the $\Lambda$CDM model
with $\Omega_{\rm m}=0.3$. 
Figure (\ref{fig:moments}) shows the ratio of the moments 
$a_{l}/\psi_{l}$ for $l=0,2$, and 4. Also shown is $a_{6}$ multiplied by a factor 
of 10 for the sake of clarity.  
To illustrate the differences 
arising from geometric and redshift distrotions,
 we assume that the mask 
structure  function, ${\cal C}_{_{\cal M}}$, is unity. 
The  ratio $|a_{0}/\psi^{0}|$ is small except at separations of $r\sim 60\hmpc$ 
where measuring the correlations is probably hard. 
The moments $a_{2}$ and 
$a_{4}$ are larger than $\psi_{2}$ and $\psi_{4}$ by a factor of $2$ and $10$, respectively, 
making 
the anisotropy pattern a very sensitive function of $\alpha$.

All moments $\int\psi^{\rm s}(\tmu)P_{l}(\tmu)\dd \tmu$ of the temperature correlation are expressed in terms 
of the functions ${\cal C}_{\cal M}$, $\xi_{_0}$, the redshift distortion parameter $\beta$
needed to express $\xi_{_l}$ in terms of $\xi_{_0}$, and the geometric
distortion parameter, $\alpha$.
A comparison between the moments 
should allow us to constrain $\alpha$, the bias parameter $\beta$, and the form  of  the mask.
As explained previously   
 the moment corresponding to $P_6$ arises only because of   geometric distortions.  Therefore, $\alpha$ can be constrained by 
 minimizing the square of this moment independent  
of the mask and redshift space distortions. Further, one can  assume parametric 
forms for $  {\cal C}_{\cal M}$ and  $\xi_{_0}$. The parameters of these forms as well as  $\alpha$ and
$\beta$ can then be obtained by fitting the moments as a function of $r$.

\subsection{Validity of the approach}
\label{sub:val}

During the early stages of reionization the filling factor, $F_{\rm f}$, of ionized regions
is small and the effect of the mask on the temperature correlations
is negligible. Further, one can  
restrict the correlation analysis to large neutral patches
extracted from 21 cm maps (figure 3  in Ciardi \& Madau 2003). 
In the intermediate and late stages of reionization ($F_{\rm f}\gsim 0.5$), it will be increasingly difficult to extract large neutral 
regions from the data (Nusser et. al. 2002). During these  stages,
the temperature correlation from the full maps will probably need to be considered. 
The method outlined    above is  capable of treating these 
stages where the mask is to be considered. 
In developing the method we have made two important assumptions related to the mask. The first is that the mask-temperature correlation is negligible. The second is that peculiar motions do not introduce any systematic 
anisotropy in the shape of the mask boundaries, i.e., the ionization fronts. Once the typical 
size of ionized regions becomes larger than the
correlation (coherence) length ($\lsim 1 \hmpc$ at $z\sim 10$) of mass distribution the mask-temperature  correlation is expected to be negligible. Figure 3 illustrates  that the 
mask-signal correlation is indeed negligible once the ionized volume is 
large enough (corresponding to $F_{\rm f} \gsim 0.13$ in the example
given in the figure). 
Let us now justify neglecting the effect of peculiar motions on the 
 mask boundaries. 
The relative displacement of the boundary is $v/(aHR)$ where  
$R $ is the typical comoving size of the ionization front, and $v$ is the typical difference in physical
peculiar velocity between two points at which a  line of sight intersects the boundary.
Also, $a=1/(1+z)$ and $H$ is the Hubble function at redshift $z$.
For simplicity consider $\Omega=1$. In this case $v=v_0 a^2 H/H_0$ where $v_0$ and $H_0$
are $v$ and $H$ at $z=0$. The relative displacement is  then $a v_0/(H_0 R)$.
Taking $H_0 R=1000$km/s which is appropriate for $F_{\rm f}\gsim 0.5$, $a=0.1$, and $v_0=100$km/s one gets a relative displacement 
of a few per cents. 
There is also an  important factor which reduces any systematic 
distortions of the ionization fronts. 
For large ionized regions the  the relative velocity $v$ can equally be  positive as much as negative. This is because 
UV ionized bubbles are  significantly larger than 
the coherence length ($\sim 1\hmpc$ at $z\sim 10$) of the dense regions near which the sources  are likely to reside.
Because of this symmetry in $v$, the mean distortions in the boundaries
of bubbles tend to cancel out.

\section{Alternative applications of the AP test}
\label{s:alternative}
So far we have considered the anisotropies of the correlations in the 
diffuse neutral component. We now consider two alternative applications 
which are also promising and are practically 
unaffected by 
foreground contamination. 
\subsection{Dense peaks and minihalos}
Consider a UV dominated reionization scenario. 
At any redshift, most of the 21 cm signal comes from the regions that have not yet been ionized.
However, the already ionized regions are likely to contain 
high density peaks in which the neutral hydrogen is dense enough to 
produce significant 21 cm emission.  
For sufficient instrumental spatial resolution these peaks would 
show  localized 21 cm emission engulfed in regions with practically  no 21 cm signature
at all. 
These peaks are particularly useful in the absence of 21 cm signal from 
diffuse neutral hydrogen towards the end of reionization.
Provided that these peaks are at least marginally resolved, the statistical anisotropy 
of their shapes can be used to constrain the cosmological parameters. This application 
is along the lines of the classic AP test (Alcock \& Paczy\'nski 1979). 
Since the peaks are compact, foreground contamination is minimized in 
this way of applying the test. 
The velocity flow  around the peaks is in the mildly non-linear regime and 
non-linear methods for modeling redshift distortions are needed. 

 The correlation function of the 
discrete distribution of the peaks could be computed. The anisotropy of this correlation could be used as a probe of geometric distortions (e.g. Ballinger, Peacock \& Heavens 1996; da \^Angela, Outram \& Shanks 2005). Here again, foreground contaminations are irrelevant since the 21 cm maps are used merely for 
locating the peaks. 
One can also consider neutral gas in minihalos in which collisions could be efficient at raising the 
spin temperature above that of the CMB (Iliev et al 2002).  They amount to significant 21 cm emission  even before reionization. Here also once 
can apply the AP test to the distribution of these
minihalos   in 21 cm maps.  
The distribution of these peaks would also be affected by redshift distortions. The signature of these distortions here is also distinguishable from that of geometric distortions. 
\subsection{The patchy mask: ionization fronts}

 In a UV radiation dominated reionization, the mask made by ionized regions is 
 patchy (see figure 3 of Ciardi \& Madau 2002).
The boundaries of these patches, i.e.,  the ionization fronts are marked by a sharp transition in 21 cm emission and 
should be possible to identify in future 21 cm maps.
The shape of the ionization fronts could be used to probe  geometric 
distortions.  
As we have seen in \S\ref{sub:val}, the effect of peculiar motions on the anisotropy of the boundaries is negligible for large enough bubbles.
Methods for identifying these boundaries in 
noisy maps observed with a given instrumental resolution must be developed.  
This is worthwhile since foreground contamination are not likely to 
affect the identification of these boundaries and also, by restricting the
application to large ionized regions, 
  redshift distortions can be made negligible.

\section{Discussion} 
\label{s:disc}

We have discussed  the  application of the Alcock-Paczy\'nski  test to 
maps of the  redshifted 21-cm emission from the era of reionization. 
Our two main conclusions are a) at $z\sim 20$ the ratio of the frequency to angular distance 
scales is sensitive to the assumed cosmological parameters and the equation 
of state of the dark energy,  and b) that geometric distortions resulting from an 
erroneous choice for the cosmological parameters are very distinct from
redshift  distortions caused by peculiar motions. Redshift distortions 
 enhance the clustering along the line of sight while geometric distortions
 simply compress the density contours either along line of sight or the
 perpendicular direction.
Judging by the ratio of the moments (see figure \ref{fig:moments}) a determination of the correlation of temperature fluctuations to an accuracy of $20\%$ should
allow a successful application of the AP test.   
Prior to the end of reionization  by UV sources 
large unionized  regions   should be easily distinguishable 
from the propagating ionized bubbles in 21 cm emission maps. 
Temperature correlation functions over scales up to several Mpcs can be 
computed reliably from these regions alone. Thus an application of the 
AP test to these correlations
completely avoid  dealing with the mask, $\cal M$. 
Nevertheless,  for the sake of completeness we have presented results including the mask.  
The prospects for measuring the cosmological 21-cm signal are good in view of 
the planed telescopes
like the Low Frequency Array \footnote{http://www.lofar.org} (LOFAR), 
the Primeval Structure Telescope \footnote{http://astrophysics.phys.cmu.edu/~jbp}
(Peterson, Pen \& We 2005), 
and the Square Kilometer Array\footnote{http://www.skatelescope.org}.
It remains to be seen how well the correlation function will be determined from data provided
by these telescopes. 

A potential obstacle in extracting any cosmological signal from 
future 21 cm maps is contamination by   bright extragalactic 
sources and Galactic synchrotron and free-free emision.
 (e.g. 
Oh \& Mack 2003; Di Matteo \etal 2002; Di Matteo, Ciardi \& Miniati 2004, Wang \etal 2005). 
 Galactic contamination contributes about 
 70\% (e.g. Wang \etal 2005). 
 This component  is believed to be a smooth function of 
 frequency. If so then it would be possible to remove its contribution 
 to the correlation function (or line of sight power spectrum) 
 of the cosmological signal on small scales. 
 The extragalactic point source contamination is more fluctuating than the Galactic
 component. But, as has been shown by Wang etal (2005), 
 the total foreground contamination can  be almost completely cleaned  from the line
 of sight power spectrum of the 21 cm emission on scales with wavenumbers
 $\ll 0.1 \rm h/Mpc$. Therefore, over a wide range of scales, foreground contaminations 
  are not expected to pose a  major problem for the application of the AP test
  and for extracting other cosmological information from 21 cm maps.
 
 We have also outlined alternative method for application of the AP test. 
 A promising method relies on identifying high density peaks in 21  cm.
 These peaks could be associated with minihalos which could lead to significant 
 emission even before the onset of reionization. They could also result 
 from neutral hydrogen  in dense regions engulfed by 
 the expanding ionized bubbles during the late stages reionization.
 Another method relies on the geometric distortions of the boundaries of 
 patchy ionized regions, i.e. the reionization fronts.  The advantage of these method is that they are 
 not affected by foreground contamination since the 21 cm maps are used merely 
 to locate the position of the peaks in the first method and the reionization
 fronts in the second. 
   
A proper testing of the methodology developed here is beyond the scope
of the current paper. Ideally one would like to use radiative transfer simulations of the  reionization era. The problem, however, is that most of the simulation
boxes are still too small for a proper statistical study of reionization. 
The largest simulation available to us is that of
Ciardi \& Madau 2003, having a cubic box of $20\hmpc$ on the side.
We intend to use this simulation to study the 
distortions in the correlations and the shapes of the ionization fronts.
An alternative to full radiative transfer simulations is the  approximate methods
developed by Benson et. al 2001 to model reionization using large N-body 
simulations.  These methods  
will enable us to study the proposed application of the AP test to simulations
of box size of $\sim 100 \hmpc$.


\section{Acknowledgment}
The author wishes to than S. Zaroubi and A.G. de Bruyn for stimulating discussions.  
This work is supported by the Research and Training Network ``The physics of the Intergalactic Medium'' set up by the European Community under the contract HPRNCT-2000-00126 and by the German Israeli Foundation for the Development of 
Research. The author is grateful to the Institute for Advanced Study (Princeton)
and the Institute of Astronomy (Cambridge) for the hospitality and support.



\end{document}